\documentclass[prl,aps,twocolumn,showpacs,floatfix]{revtex4}
\usepackage{graphicx}

\begin{document}

The work by Finn {\em et al} further investigating our system is indeed 
relevant. Their results, particularly the complete absence of chaos for 
the $\Gamma = 0.3$ case, are somewhat surprising and inconsistent with 
our results. 

Our published results reported analysis on the in-principle experimentally 
accessible time-series. For $\Gamma = 0.125$ Poincar\'e sections indicate 
a chaotic attractor at $\beta = 0.01$, which is altered but persists for 
$\beta = 0.3$ and disappears for $\beta = 1$. 
For $\Gamma = 0.3$, Poincar\'e sections show no chaos at $\beta = 0.01$, 
an attractor for $\beta = 0.3$, which disappears for $\beta = 1$. 
The power spectra for $\langle \hat X(t) \rangle$ for all six cases agree
with the above.
Further, the $\beta = 0.3$ results look extremely similar for the two 
$\Gamma$ cases. 

\begin{figure}[htbp]
\centerline{\includegraphics[width=8.3cm,height=10.8cm,clip]
{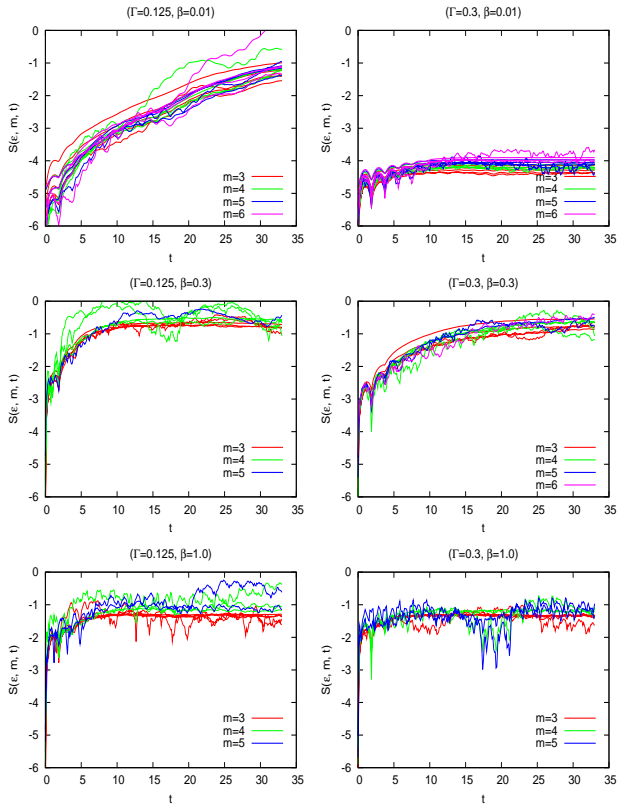}}
\caption{
Each plot shows the divergence $S(\epsilon,m,t)=\ln(\Delta(t))$ of nearby 
points in the reconstructed phase space for a given $\Gamma,\beta$ pair. 
There are several curves in each plot with different delay embedding dimension
$m$; for each $m$ there are curves corresponding to several different 
neighborhood radii $\epsilon$ in the reconstructed phase space. 
Exponential growth appears as linearity before the trajectories reach 
saturation; the slope is proportional to $\lambda$. See
Ref.~\cite{tisean} for details on the technique.
}
\label{figsix}
\end{figure}
We have since used the TISEAN package\cite{tisean}, performing phase-space 
delay reconstruction with $\langle \hat X(t) \rangle$ to obtain $\lambda$. 
We see qualitative agreement with our previous results (see Fig.~(1)). 
We estimate $\lambda$ for respective $(\Gamma, \beta)$ pairs 
(approximately, since they derive from finding the slope of the straight 
line parts of these curves) as: $(0.125,0.01) \approx 0.1, (0.125,0.3) 
\approx 0.16, (0.125,1.0) < 0.05, (0.3,0.01) < 0.03, (0.3,0.3) \approx 0.13, 
(0.3,1.0) < 0.05$. In short, this agrees with our previous conclusions about 
where chaos exists.  Interestingly, using $\lambda$, the transition 
from quantum to classical behavior appears to be non-monotonic for {\em both} 
instances of $\Gamma$.

Our three methods of analysis (Poincar\'e sections, power spectra, and 
time-series Lyapunov exponents) are all consistent with each other, and 
consistent with our physical understanding of how the chaos emerges 
and/or is swamped by quantum effects. Finn {\em et al}'s calculation is 
inconsistent with this for the one `mesoscopic' case of 
$(\Gamma,\beta) =(0.3,0.3)$ and we are particularly surprised that 
their results for the $(0.125,0.3)$ and $(0.3,0.3)$ 
cases are so different. As noted by an anonymous referee, it is possible
that the chaos is a finite-time effect in a system where the 
infinite-time limit is non-chaotic. Of course, finite-time behavior 
is also physically important, and could be of greater physical relevance 
than the mathematical infinite-time limit in real experimental applications.

We expect that understanding the source of this 
difference --- provided it is not due to technical errors --- will reveal 
something deeper about the physics, or about the difference between the 
methods of analysis. Behind the immediate questions about the behavior of 
this model system stands the larger fundamental question of whether quantum 
corrections always regularize and suppress chaotic dynamics. We believe that 
this, while often true, is not universal.  For the QSD equations (or equivalent 
stochastic Schrodinger equations) it is extremely unlikely that such a 
highly nonlinear equation has a priori a monotonic parameter landscape. 
Our perspective is supported, for example, by Bhattacharyya 
{\em et al} \cite{bhatt}. It is only a matter of more systematic 
investigation to find other such counter-examples to the folklore.

Kyle Kingsbury$^{(a)}$

Chris Amey$^{(a)}$

Arie Kapulkin$^{(b)}$

Arjendu Pattanayak$^{(a)}$

{(a) Department of Physics and Astronomy, \\
Carleton College, Northfield, Minnesota 55057
\\
(b) 128 Rockwood Cr, Thornhill, Ont L4J 7W1 Canada}

\end{document}